# Building A High Performance Parallel File System Using Grid Datafarm and ROOT I/O

Y. Morita, H. Sato, Y. Watase

*KEK, Tsukuba, Ibaraki 305-0801, JAPAN*

O. Tatebe, S. Sekiguchi

*AIST, Tsukuba, Ibaraki 305-8568, JAPAN*

S. Matsuoka

*Tokyo Institute of Technology, Meguro, Tokyo 152-8552, JAPAN*

N. Soda

*Software Research Associates, Inc., Naka, Nagoya, 460-0003, JAPAN*

A. Dell'Acqua

*CERN, CH-1211, Geneve 23, Switzerland*

Sheer amount of petabyte scale data foreseen in the LHC experiments require a careful consideration of the persistency design and the system design in the world-wide distributed computing. Event parallelism of the HENP data analysis enables us to take maximum advantage of the high performance cluster computing and networking when we keep the parallelism both in the data processing phase, in the data management phase, and in the data transfer phase. A modular architecture of FADS/Goofy, a versatile detector simulation framework for Geant4, enables an easy choice of plug-in facilities for persistency technologies such as Objectivity/DB and ROOT I/O. The framework is designed to work naturally with the parallel file system of Grid Datafarm (Gfarm). FADS/Goofy is proven to generate $10^6$ Geant4-simulated Atlas Mockup events using a 512 CPU PC cluster. The data in ROOT I/O files is replicated using Gfarm file system. The histogram information is collected from the distributed ROOT files. During the data replication it has been demonstrated to achieve more than 2.3 Gbps data transfer rate between the PC clusters over seven participating PC clusters in the United States and in Japan.

## 1. INTRODUCTION

MONARC collaboration has proposed a multi-tier and hierarchical world-wide networks of computing centers for widely distributed data analysis for the LHC experiments [1]. Grid technology is foreseen to provide a secure, scalable and industrial standard distributed computing environment [2]. In such environment, efficient and well-designed data replication and job scheduling algorithms are necessary by carefully simulating the system behavior [3].

Typical data analysis in high energy physics experiments are based on the data unit of events. To take maximum advantage of the event parallelism to achieve highly scalable CPU and I/O performance, both in the local area environment and in the world-wide environment, we have proposed and developed Grid Datafarm architecture [4]. A raw I/O benchmark with the parallel file system has achieved 1.97 GB/s with 64 nodes [5]. Combined with the Geant4 detector simulation framework FADS/Goofy, a good scalability has been demonstrated with Objectivity/DB as an I/O module [6].

In the view point of software architecture, experiments' software suit must be properly insulated from the technology choices of the underlying hardware and software resources. Discussion towards the LHC-common solution for the lower level software suits [7] made it clear that the framework for the high energy physics data analysis must be modular, robust and lightweight, and yet it must work with a highly scalable computing environment.

In this paper we report the software development of the modular architecture of the FADS/Goofy I/O framework, and the performance tests of the simulated detector hit information using ROOT I/O.

## 2. SOFTWARE SUIT

### 2.1. FADS/Goofy

FADS/Goofy is a light weight, object-oriented framework for the detector simulation using Geant4 [8,9]. It consists of a small, autonomous executable *goofy* and a set of framework service shared libraries. Late binding mechanism of C++ allows *goofy* to load the service modules at runtime, thus expanding the functionality of the framework without re-linking *goofy*. Modular detector construction factories allow users to build the Geant4 geometry with XML detector description, NOVA/MySQL database [10], or the native Geant4 geometry classes. Similarly, histogramming and graphic display libraries can be used with the corresponding service modules for HBOOK [11] and ROOT [12].

User detector code can also be built using the same late binding mechanism, and it provides an easy to build, quick turn-around and modular detector code development for a very complex detector geometry such as ATLAS.

Once the detector code is validated, the same set of shared libraries can also be used in the ATLAS main analysis framework, called ATHENA [13].

**TUDT010**

## 2.2. ROOT I/O

ROOT is a powerful and flexible statistical data analysis tool [11], and it can also be used as a simple object streaming package with ROOT TTree object.

We have developed the persistency module of FADS/Goofy in such a way that the object I/O is controlled through the transaction managers of user defined objects. Transient objects are not affected by changing the different I/O service modules such as for Objectivity/DB and ROOT I/O.

For each sub detector, user has to provide a Geant4 hit class and the I/O property definition file called *.rootio. A perl script *fadsrootio.pl* then automatically generates the necessary adapter classes for the ROOT I/O (fig. 1).

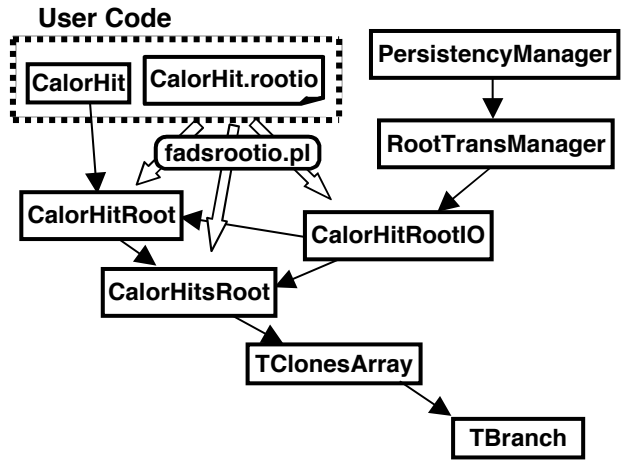

Figure 1: *fadsrootio.pl* generates the ROOT I/O module classes based on the user defined *.rootio* file.

At runtime, hit collections from each sub detector are stored into a ROOT Tree Branch as a TClonesArray at each event.

Figure 2 shows an example of the user defined *.rootio* file. It resembles to writing a persistent object description file for storing the object data member with a constructor method and retrieving with a make_transient method. User can utilize predefined macros such as *@float@* and user defined macro *@class_name@* to make the description universal and portable for other part of the detector, or for the other I/O service module of FADS/Goofy.

This persistency scheme has been fed back to the Geant4 5.0 persistency examples as *g4rootio.pl*.

## 2.3. Gfarm

The Grid Datafarm (Gfarm) is architecture for petascale data-intensive computing on the Grid [4]. The model specifically targets applications where data primarily consists of a set of records or objects which are analyzed independently. Gfarm takes advantage of the data access locality to achieve a scalable I/O bandwidth using an enhanced parallel file system integrated with process scheduling and file distribution. It provides a global, Grid-enabled, fault-tolerant parallel file system whose I/O bandwidth scales to the TB/s range, and which incorporates fast file transfer techniques and wide-area replica management.

```
set class_name Pers01CalorHit
set collection_class Pers01CalorHitsCollection
set collection_base_class G4VHitsCollection
set sdet_name Pers01CalorHit
set array_io_base VPHitsCollectionIO
set catalog HCIOentryT
set global_declaration
  class @class_name@;    // forward declaration
..
set add_header_src
 @class_name@.hh
 G4ThreeVector.hh
 G4RotationMatrix.hh
..
set member
 @float@  EdepAbs;
 @float@  EdepGap;
 @float@  TrackLengthAbs;
 @float@  TrackLengthGap;
..
set constructor
 @class_root@(@class_name@* hit)
 {
   // copy data members of transient hit
   EdepAbs = hit->GetEdepAbs();
   EdepGap = hit->GetEdepGap();
   TrackLengthAbs = hit->GetTrakAbs();
   TrackLengthGap = hit->GetTrakGap();
 }
..
set method
 @class_name@* @make_transient@()
 {
   // create a transient class
   @class_name@* hit = new @class_name@();

   hit->AddAbs(EdepAbs, TrackLengthAbs);
   hit->AddGap(EdepGap, TrackLengthGap);

   return hit;
 }
..
```

Figure 2: Example of the user defined *.rootio* file.

For the case of FADS/Goofy, the ROOT I/O file is a collection of events which in turn are collections of TClonesArray of the sub detector hits which can be stored and analyzed independently. This data model makes it natural for take maximum advantage of the Gfarm parallel file system.

FADS/Goofy jobs submitted to the Gfarm are distributed to the cluster nodes, and the files generated at each node are registered as a single local file into the metadata catalog. Users can replicate the logical files for their analysis, fault-tolerant backups, or for an efficient file transfer over wide-area network with parallel transfer.

## 3. PERFORMANCE EVALUATION

To test the scalability of the Gfarm-based ROOT I/O performance with FADS/Goofy, 256 nodes dual Athlon MP 1900+ (1.6GHz) PC cluster Presto III, at Tokyo

Institute of Technology, was used. Each node is equipped with 756MB memory and an interface to Myrinet 2K. Total capacity of the IDE disks is 100TB.

ROOT I/O persistency module for a simple calorimeter hit class was used for the I/O benchmark. Four float values in Figure 2 were filled with gaussian random numbers, and stored and retrieved with the Gfarm parallel file system. Hits were repeated 1,000 times in each event. Typical compression factor with ROOT I/O was 3.5 and average event size was 7.5KB.

Figure 3 shows the aggregated ROOT I/O performance with the number of parallel nodes. A good scalability was obtained up to 32 nodes in the case of reading. By increasing nodes, aggregated throughput was dragged down by the inclusions of several slow nodes.

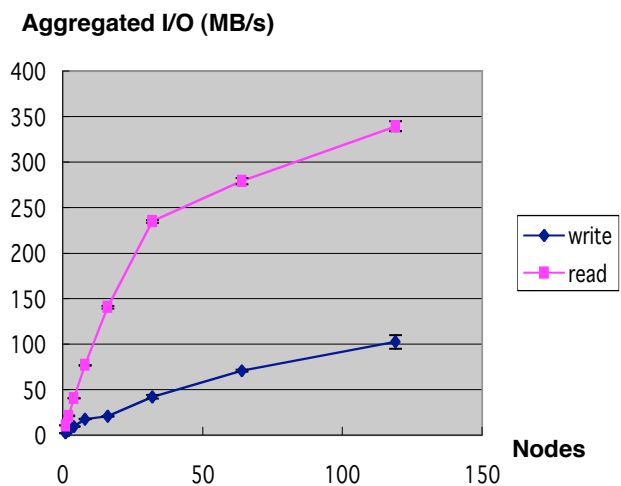

Figure 3: Aggregated throughput of FADS/Goofy ROOT I/O module with the parallel file system of Gfarm.

Detail inspection of the dragging nodes revealed various reasons of different performance behaviors. In one case there were some additional processes consuming the memory of the node, such as sshd. In other cases the disk utilization and the fragmentation was high. In some nodes disks were replaced with new ones, and the throughput was improved by factor 4 or more. In other words, running the parallel I/O jobs with Gfarm is a good screening test for those non-performing nodes.

It should be noted that the variance of the node performance is dominated in the real PC cluster environment, and every effort should be made to normalize the participating nodes for achieving the high throughput computing. The architectural limit of the Gfarm parallel file system overhead is not visible in this measurement.

## 4. RELATED WORKS

The ROOT I/O persistency module was also used to produce the full simulated ATLAS events with the Higgs to four muon Monte Carlo generated events for the robustness evaluation. Full detector simulation is a CPU-bound job and a good parallelism was obtained for generating one million events in two days, with 500 CPUs of Presto III PC cluster.

The simulated data was then replicated to the other PC clusters from Tokyo Institute of Technology to AIST through KEK with NII SuperSINET and IMnet. It was then replicated to the United States with APAN/TransPAC network, to participate in the Bandwidth Challenge of SC2002. Aggregated file replication rate of 2.3 Gbps between the SC2002 site and the other participating sites. For the file replication over the Pacific, 741 Mbps transfer rate has been achieved out of 893 Mbps link [14].

## 5. CONCLUDING REMARKS

A modular and versatile persistency framework has been developed for FADS/Goofy. Its design allows the user code to be isolated from the persistency technology choice. Adapter classes for ROOT I/O can be generated automatically with a perl script. Same user code can be utilized for the Objectivity/DB or other persistency packages.

A standalone performance test of the ROOT I/O persistency module has been conducted on the Gfarm parallel file system using PrestoIII, a PC cluster of Tokyo Institute of Technology. A good scalability was obtained up to a few tens of nodes, where differences in various types of node performance in the real environment become apparent. Aggregated throughput of the parallel file system is sensitive to the performance of the few dragging nodes. The architectural limit of the Gfarm is not visible in this measurement.

Once the event data files are generated on the Gfarm parallel file system, it was demonstrated to be very efficient in replicating the files with Gfarm file replication mechanism over wide area networks.

## Acknowledgments

The authors wish to thank NII, the National Institute of Informatics, and the KEK Network Group for their support on high speed network connectivity in Japanese academic institutions and to overseas connections. The authors are also grateful for the continuous support of PrestoIII PC cluster by the staff and students of Matsuoka Lab., especially Mr. K. Shirose and Mr. Y. Takamiya of Tokyo Institute of Technology. Mr. Tomeda's work at Okayama University on building and checking the detector geometry of the silicon tracker for the ATLAS inner detector was invaluable to this work.

Work supported in part by Ministry of Education, Culture, Sports, Science and Technology, Kakenhi Tokutei Ryoiki (2) No. 13224034.